\documentclass[sigconf,anonymous=false]{acmart}
\usepackage[utf8]{inputenc}
\usepackage[T1]{fontenc}    
\usepackage{appendix,titlepic}

\usepackage{booktabs,tabularx}
\usepackage{amsxtra}

\usepackage{wrapfig, blindtext}

\usepackage{caption}

\usepackage{makecell}

\usepackage{enumitem}
\usepackage{lscape}
\usepackage{rotating}
\usepackage{epstopdf}
\usepackage[flushleft]{threeparttable}
\usepackage{etoolbox}
\usepackage{nomencl}
\usepackage{amsthm}
\usepackage{supertabular}
\usepackage{amsfonts}
\usepackage[dvips]{epsfig}
\usepackage{latexsym}
\usepackage{amsmath}
\usepackage{bbm}
\usepackage{verbatim}
\usepackage[ruled,linesnumbered,boxed,commentsnumbered]{algorithm2e}
\usepackage{bm}
\usepackage{tikz}
\usepackage{mathtools}
\usepackage{xcolor}

\AtBeginDocument{%
  \providecommand\BibTeX{{%
    \normalfont B\kern-0.5em{\scshape i\kern-0.25em b}\kern-0.8em\TeX}}}

\copyrightyear{2023} 
\acmYear{2023} 
\setcopyright{acmlicensed}\acmConference[e-Energy '23]{The 14th ACM International Conference on Future Energy Systems}{June 20--23, 2023}{Orlando, FL, USA}
\acmBooktitle{The 14th ACM International Conference on Future Energy Systems (e-Energy '23), June 20--23, 2023, Orlando, FL, USA}
\acmPrice{15.00}
\acmDOI{10.1145/3575813.3576871}
\acmISBN{979-8-4007-0032-3/23/06}

\usepackage{lineno}
\modulolinenumbers[5]
\usepackage{url}
\usepackage{amsmath}

\usepackage{multirow}
\usepackage{comment}

\usepackage{xcolor}

\begin{document}
\title{Model-Free Approach to Fair Solar PV Curtailment Using Reinforcement Learning}


\author{Zhuo Wei, Frits de Nijs, Jinhao Li, Hao Wang}
\authornote{Corresponding author: \url{hao.wang2@monash.edu} (Hao Wang).}
\affiliation{
  \department{Department of Data Science and AI}
  \institution{Monash University}
  \city{Melbourne}
  \state{Victoria}
  \country{Australia}
  \postcode{3800}
}


\begin{abstract}
The rapid adoption of residential solar photovoltaics (PV) has resulted in regular overvoltage events, due to correlated reverse power flows. Currently, PV inverters prevent damage to electronics by curtailing energy production in response to overvoltage. However, this disproportionately affects households at the far end of the feeder, leading to an unfair allocation of the potential value of energy produced. Globally optimizing for fair curtailment requires accurate feeder parameters, which are often unknown. 
This paper investigates reinforcement learning, which gradually optimizes a fair PV curtailment strategy by interacting with the system. We evaluate six fairness metrics on how well they can be learned compared to an optimal solution oracle. We show that all definitions permit efficient learning, suggesting that reinforcement learning is a promising approach to achieving both safe and fair PV coordination.
\end{abstract}




\maketitle

\section{Introduction}\label{1}
Distributed energy resources owned by households are growing in popularity. 
Installed capacity of photovoltaic (PV) systems has grown at an average annual rate of 49\% over the last decade~\cite{sevilla2018techno}, with the global PV capacity expected to more than double in the next decade~\cite{o2020too}. However, PV systems can result in significant challenges to the local grid because of the midday PV power injection and reverse power flow \cite{zhang2021high}. One of the most severe problems is that the massive increase in PV systems can lead to significant excess generation and overvoltage problems on distribution feeders~\cite{denholm2015overgeneration}.

Active power curtailment has been proposed as an important solution to the technical and economic challenges associated with PV power generation \cite{sevilla2018techno}. Inverter reactive power compensation as a method of PV curtailment can be an effective solution to solve overvoltage problems. However, this method often results in more PV curtailment for households at the end of the feeder \cite{gebbran2021fair}. 
Distribution network operators urgently need a fair PV curtailment solution. For example, households with PV in Australia currently lose a total value of $1.2$-$4.5$ million Australian dollars per year due to unfair PV curtailment \cite{stringer2021fair}. Also, according to \citet{carley2020justice}, the transition to low-carbon energy sources, including PV generation, can perpetuate existing energy inequity if the fairness problem cannot be solved. Therefore, the risk of unfairness may become a barrier to entry for late adopters of PV panels, slowing down the overall transition to a renewable grid and net zero.

Although fair PV curtailment solutions exist, practical technical challenges remain unsolved. \citet{gebbran2021fair} used an optimization-based method to achieve fair PV curtailment, assuming known distribution network parameters and accurate load forecasts. However, distribution network visibility is a challenging problem, and network parameters are often not known \cite{Weckx2012,li2022online}. In particular, distribution network operators may obtain incorrect grid parameters due to insufficient data updates \cite{christakou2017voltage}. A recent study by \citet{yeh2022robust} investigated voltage control with an unknown grid topology to ensure stability, but without a fairness objective. Furthermore, globally fair curtailment of solar panels also requires accurate predictions of future generations and demand to be able to ensure fairness over the full day.

We propose to approach the problem of fair PV curtailment using reinforcement learning (RL), for RL can be used in real-time and does not require exact physical models of the distribution system \cite{arwa2020reinforcement,feng2021taxonomical}. Therefore, RL particularly suits the PV curtailment problem, in which the distribution network visibility is low. We train an RL algorithm, which learns a fair PV curtailment policy from simulated control on historical data, and continuous interactions with the distribution network environment. The learned control policy aims to minimize overall electricity costs while fairly curtailing households' PV generation to keep their voltages in a safe range. 

The contributions of this paper are as follows.
\begin{itemize}
    \item We propose a new approach for fair PV curtailment using RL. Our approach does not require any knowledge of the distribution network parameters but interacts with the network and households to learn a fair PV curtailment strategy.
    \item We specify three fairness definitions for PV curtailment over two different periods, instant and accumulative over a day, resulting in six cases. These six cases are demonstrated to be effective in learning fair PV curtailment strategies.
    \item We compare the RL solutions to optimization solutions which rely on an accurate network model by characterizing the trade-off between electricity cost reduction and fairness.
\end{itemize}

\section{Model}\label{3}
To use RL methods to minimize households' electricity costs and ensure voltage safety and fair PV curtailment, we need to build an environment for a distribution network, including voltage constraints and fairness definitions for PV curtailment. 

\subsection{Network Model}\label{subsec:net}
We introduce a linear feeder network~\cite{peng2016distributed} as the environment to capture the interactions between households and voltage changes on the network caused by PV injection, a simplification which allows us to calculate and compare with the optimal curtailment. The RL algorithm presented in Section~\ref{subsec:pd} will interact with the developed environment to learn a fair PV curtailment strategy.

\begin{figure}[t]
    \centering
    \input{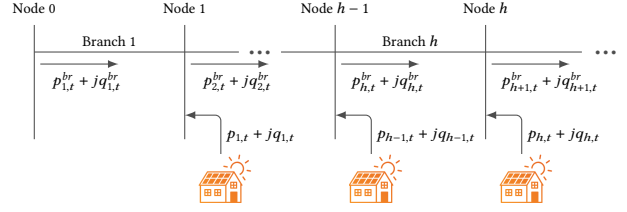}
    \caption{The structure of the network.}
    \label{f1}
\end{figure}

Specifically, our feeder network, shown in Fig.~\ref{f1}, captures the dependency of voltage changes with power import and export on each node. We denote $p^{br}_{h,t}$ and $q^{br}_{h,t}$ as the active and reactive power at time $t$ on branch $h$, and $j$ is the imaginary unit. 
For each household $h$, we denote $p_{h,t}$ and $q_{h,t}$ as the active power and reactive power exchanged between the household and the grid at time $t$, respectively. When $p_{h,t} < 0$, household $h$ imports power from the grid; otherwise when $p_{h,t} > 0$, the household exports power to the grid. Similarly for reactive power, $q_{h,t} < 0$ indicates absorption, and $q_{h,t} > 0$ indicates release of reactive power. According to \citet{zhong2018topology}, the power flows in Fig. \ref{f1} are described by
\begin{align}
p^{\textit{br}}_{h+1,t} &= p^{\textit{br}}_{h,t} + p_{h,t}, \\
q^{\textit{br}}_{h+1,t} &= q^{\textit{br}}_{h,t} + q_{h,t}, \\
v_{h+1,t} &= v_{h,t} - (r_{h+1,t}\,p^{\textit{br}}_{h+1,t} + x_{h+1,t}\,q^{\textit{br}}_{h+1,t}) / v_{0,t},
\end{align}
where $v_{h,t}$ is the voltage of node $h$ at $t$, $r_{h,t}$ and $x_{h,t}$ are the resistance and reactance of branch $h$, and node $0$ connects to the grid.

\subsection{Electricity Cost and Feed-in Tariff}\label{subsec:pd}
We aim to minimize households' electricity costs, which  means maximizing households' profits from exporting PV. Therefore, the objective of the problem is described as

\begin{equation}
    \min \sum_{h \in \mathcal{H}} \sum_{t \in \mathcal{T}} {c(t) \times (p^L_{h,t} - p^G_{h,t} \times (1 - \alpha^C_{h,t}))},
\end{equation}
where $\mathcal{H}$ is the set of households, and $\mathcal{T}$ is the horizon in which fair PV curtailment is implemented. The electricity price is denoted as $c(t)$, as it is often a time-of-use (ToU) price \cite{gebbran2021fair}. For household $h$, PV generation at $t$ is $p^G_{h,t}$, and $p^L_{h,t}$ denotes household $h$'s load at $t$. For each household, a certain percentage of PV generation curtailed is a decision variable, which is denoted as $\alpha^C_{h,t}$. Thus, the net power exchange between household $h$ and the grid $p_{h,t}$ is expressed as
\begin{equation}
    p_{h,t} = P^G_{h,t} \times (1 - \alpha^C_{h,t}) - p^L_{h,t}.
\end{equation}

\subsection{Voltage Control and Constraint}\label{subsec:safe}
The safe range of the voltage is often set between 0.95 and 1.05 per unit (p.u.) \cite{peng2016distributed}. Thus, the voltage constraints can be expressed as
\begin{equation}
    0.95 \leq v_{h,t}\text{ (p.u)} \leq 1.05, \forall \text{ }h \in \mathcal{H}, t \in \mathcal{T},
\end{equation}
where $v_{h,t}$ is the voltage of node $h$ at $t$.

PV inverters are used for voltage control to ensure that the voltages are within the safe range. Specifically, PV inverters support Volt-Var Control (VVC) \cite{gebbran2021fair}, in which voltages are controlled by injecting or absorbing reactive power in response to voltage. The relationship between $q_{h,t}$ and $v_{h,t}$ can be expressed as
\begin{equation}
\begin{aligned}
& q_{h,t} = \overline{q} \times \mathbbm{1}(v_{h,t} \leq V_1) + \overline{q}\frac{V_2 - v_{h,t}}{V_2 - V_1} \times \mathbbm{1}(V_1 < v_{h,t} < V_2) \\& - \overline{q}\frac{v_{h,t} - V_3}{V_4 - V_3} \times \mathbbm{1}(V_3 < v_{h,t} < V_4) - \overline{q} \times \mathbbm{1}(v_{h,t} \geq V_4),
\end{aligned}
\end{equation}
where $\mathbbm{1}(\cdot)$ is the indicator function, $\overline{q}$ is the maximum reactive power output, which is normally defined as a percentage of the maximum capacity of the PV inverter in each household. The values of voltage thresholds $V_1$ to $V_4$ depend on different product standards.

Also, due to the use of PV inverters, it is necessary to consider the capacity of the inverters, whose relationship with the PV curtailment can be expressed as
\begin{equation}
    \overline{s}^2 \geq (p^G_{h,t} \times (1 - \alpha^C_{h,t}))^2 + q_{h,t}^2, \forall \text{ }h \in \mathcal{H}, t \in \mathcal{T},
\end{equation}
where $\overline{s}$ is the maximum capacity of PV inverters.

\subsection{Definition of Fairness}\label{subsec:fair}
Aiming to curtail PV fairly, we consider three definitions of fairness. According to \cite{carley2020justice}, the core principle of energy fairness is to ensure that certain groups do not disproportionately share the burden or fail to receive the benefits. In our problem, fair PV curtailment means that no households suffer from more curtailment or low PV export compared with other households. We follow the approach of \cite{gebbran2021fair}, who defines three fairness metrics:
\begin{enumerate}
    \item Definition $1$: egalitarian PV curtailment;
    \item Definition $2$: proportional PV curtailment;
    \item Definition $3$: egalitarian PV output.
\end{enumerate}

In Definition $1$, the PV curtailment of each household does not exceed a specific value. In Definition $2$, the ratio of PV curtailment to the maximum available export power for each household does not exceed an upper limit. In Definition $3$, the PV exports of households are not below the lower bound. The differences between the three definitions are illustrated in Appendix \ref{ap1}.

The work of \cite{gebbran2021fair} proposed fairness definitions that should be optimized strictly, where fairness is maximized every timestep. We consider a relaxed \emph{accumulative} case, which allows the controller to gradually equalize fairness over the full length of a day. For Definition $1$, fair PV curtailment means that the \emph{maximum} amount of PV curtailment needs to be minimized. Therefore, in these two cases, the fairness penalty due to PV curtailment becomes
\begin{align}
    \boldsymbol{\alpha}^C_t &= \bigl\{\alpha^C_{h,t} \mid \forall h \in \mathcal{H}\bigr\}, \\
    f^{\textit{ins}}_{1}(\boldsymbol{\alpha}^C_t) &=  \max_{h \in \mathcal{H}}\left(\alpha^C_{h,t}\cdot p^G_{h,t}\right), \\
    f^{\textit{acc}}_{1}(\boldsymbol{\alpha}^C_T) &=  \max_{h \in \mathcal{H}}\left(\textstyle\sum_{t=0}^T \bigl( \alpha^C_{h,t}\cdot p^G_{h,t}\cdot \Delta\bigr)\right),
\end{align}
where $T$ is the control horizon of one day in thirty-minute steps. To show accumulative quantities in kWh, step size $\Delta = 30/60$, which means each half hour.

For the Definition $2$, the ratio of PV curtailment to the maximum available export power needs to be minimized. Moreover, this definition only applies to the case where the PV generation of households is more than their power consumption. Thus, the fairness penalty in the two cases can be expressed as
\begin{align}
f^{\textit{ins}}_{2}\bigl(\boldsymbol{\alpha}^C_t\bigr)
  &= \max_{h \in \mathcal{H}} 
      \left(
      \frac
        {\alpha^C_{h,t} \cdot p^G_{h,t}}
        {\max\bigl(\epsilon, p^G_{h,t} - p^L_{h,t}\bigr)}
      \right),
\\
f^{\textit{acc}}_{2}\bigl(\boldsymbol{\alpha}^C_T\bigr)
  &= \max_{h \in \mathcal{H}}
      \left(
      \frac
        {\sum_{t=0}^T \bigl(\alpha^C_{h,t}\cdot p^G_{h,t} \cdot \Delta\bigr)}
        {\max\bigl(
            \epsilon,
            e^E_h
         \bigr)
        }
      \right),
\\
e^E_h &= \sum_{t = 0}^T((p^G_{h,t} - p^L_{h,t})\cdot \Delta),
\end{align}
where $e^E_h$ is the total energy export in kWh, and $\epsilon > 0$ a small value.

For Definition $3$, the PV exports of households need to be maximized. Same as Definition $2$, this definition only applies to households where PV generation is larger than their daily consumption. Therefore, the fairness penalty in the two cases can be expressed as
\begin{align}
f^{\textit{ins}}_{3}\bigl(\boldsymbol{\alpha}^C_t\bigr)
    &= -\min_{h \in \mathcal{H}}\Bigl(
     \begin{aligned}[t]
	  &\mathbbm{1}(\alpha^C_{h,t} > 0) \cdot ((1 - \alpha^C_{h,t}) \cdot p^G_{h,t} - p^L_{h,t}) \\ 
      \mathllap{{}+{}}&\mathbbm{1}(\alpha^C_{h,t} = 0) \cdot \max_{h \in \mathcal{H}}\bigl(p^G_{h,t} - p^L_{h,t}\bigr)\Bigr),
     \end{aligned}\\
f^{\textit{acc}}_{3}\bigl(\boldsymbol{\alpha}^C_T\bigr)
    &= -\min_{h \in \mathcal{H}}\Bigl(
	 \begin{aligned}[t]
	  \mathbbm{1}(\textstyle\sum_{t=0}^T \alpha^C_{h,t} > 0) \cdot (\sum_{t=0}^T((1 - \alpha^C_{h,t}) \:\:\\
    \mathllap{{}\cdot p^G_{h,t} - p^L_{h,t})\cdot \Delta) + {}}\mathbbm{1}(\textstyle\sum_{t=0}^T \alpha^C_{h,t} = 0) \cdot \displaystyle\max_{h \in \mathcal{H}}(e^E_h)\Bigr),&
     \end{aligned}
\end{align}
where the negative sign means that more PV exports will result in less penalty, thus maximizing PV exports.

\section{Methodology}\label{4}
We use reinforcement learning (RL) to optimize a fair and safe control policy for PV curtailment. To do so, we first define a Markov Decision Process (MDP) on the single-feeder control problem, and train a controller with Soft Actor Critic~\cite{haarnoja2018}.

\subsubsection*{MDP model}\label{subsec:mdp}
An MDP is a discrete-time model defined by its state space~$\mathcal{S}$, action space~$\mathcal{A}$, transition function~$T(s,a) \to s'$ and reward function~$R(s,a) \to \mathbb{R}$, with the objective to find a control policy~$\mu$ that maximises the expected discounted cumulative reward, $\max_{\mu} \mathbb{E}\left[\sum_t \gamma^t R(s_t, a_t) \mid s_{t+1} \sim T(s_t, a_t), a_t \sim \mu(s_t) \right]$.

The state space of our agent depends on the definition of fairness we consider. The instantaneous case state~$\boldsymbol{s}^{\textit{ins}}_t$ contains the previous time's measured load, generation, and voltage, while the accumulative case~$\boldsymbol{s}^{\textit{acc}}_t$ also includes the total curtailment up to $t$
\begin{align}
\boldsymbol{s}^{\textit{ins}}_t &= \left(p_{h,t-1}, p^L_{h,t}, p^G_{h,t}, v_{h,t-1} \mid \forall h \in \mathcal{H}\right), \\
\boldsymbol{s}^{\textit{acc}}_t &= \bigl(p_{h,t-1}, p^L_{h,t}, p^G_{h,t}, v_{h,t-1}, \sum_{\tau=0}^{t-1} \alpha^C_{h,\tau} \cdot p^G_{h,\tau} \mid \forall h \in \mathcal{H}\bigr).
\end{align}

Action space $\mathcal{A}$ includes all possible actions $\boldsymbol{a}$ that the agent can perform in the environment. In this paper, one central agent decides how much PV production to curtail for all households. Thus, the action $\boldsymbol{a}_t$ is the proportion of curtailed power at moment~$t$,
\begin{equation}
\boldsymbol{a}_t = \{\alpha^C_{h,t}, \forall \text{ }h \in \mathcal{H} \}.
\end{equation}

Our objective is to maximize households’ profits from selling PV, while ensuring fair PV curtailment to maintain voltage safety. To maintain the voltage within the safe range, we set how much the voltage exceeds the safe range $[0.95,1.05]$ as the penalty term $g(v_{h,t})$, which can be expressed as
\begin{equation}
g(v_{h,t}) =
   \max{\left(0.95 - v_{h,t}, 0\right)}
 + \max{\left(v_{h,t} - 1.05, 0\right)}.
\end{equation}

For the fairness penalty caused by PV curtailment, we let $f(\boldsymbol{\alpha}^C_t)$ denote the general form of fairness penalty, representing all six cases, such as $f^{\textit{ins}}_{1}(\boldsymbol{\alpha}^C_t)$, $f^{\textit{acc}}_{1}(\boldsymbol{\alpha}^C_T)$, $f^{\textit{ins}}_{2}\bigl(\boldsymbol{\alpha}^C_t\bigr)$, $f^{\textit{acc}}_{2}\bigl(\boldsymbol{\alpha}^C_T\bigr)$, $f^{\textit{ins}}_{3}\bigl(\boldsymbol{\alpha}^C_t\bigr)$, and $f^{\textit{acc}}_{3}\bigl(\boldsymbol{\alpha}^C_T\bigr)$.

Therefore, the reward $r_t$ is expressed as
\begin{equation}
\begin{aligned}
r_t = \sum_{h \in \mathcal{H}} &\left[(1 - \alpha^C_{h,t}) - w_1 g(v_{h,t})\right] - w_2 f(\boldsymbol{\alpha}^C_t),
\end{aligned}
\label{eq:reward}
\end{equation}
where $w_1$ and $w_2$ are weights. For accumulative cases, the fairness penalty only applies once per episode, at the last step in the episode.

\subsubsection*{Learning algorithm}\label{subsec:SAC}
We use the SAC method \cite{haarnoja2018} to solve the fair PV curtailment problem. More specifically, we set a distribution network as the environment required by the RL model. Then we train an agent by the SAC algorithm to be responsible for the PV curtailment of all households. We use this agent to curtail PV generation for all households and simulate the impact of PV curtailment on the voltage and cost of electricity through the network model. Thus, the agent can interact with the environment and learn a reasonable PV curtailment policy. For our hyper-parameters, we set the discount factor to $0.99$, learning rate to $0.0003$, and batch size to $64$. We set $80\%$ of the data source as the training set and $20\%$ as the test set. The sampling is from 0:00 midnight each day. The training episodes were set to $24$ hours.

\section{Case Study}\label{5}
Preliminary experimental results are implemented in a network containing $10$ households, using data from \cite{ratnam2017residential}. We use half-hourly historical data for one year ($17,520$ data points per panel) as our training set. Also, we train the model for $3 \times 10^4$ simulated timesteps on this dataset. The learning curves of our model are shown in Appendix \ref{ap3}. We use Australian dollars (\$) to show monetary values. According to \cite{gebbran2021fair}, the time-of-use electricity prices are $0.12$ \$/kWh for off-peak (23:00--7:00), $0.52$ \$/kWh for peak (14:00--19:30), $0.22$ \$/kWh for other time periods. The feed-in tariff is $0.1$ \$/kWh throughout the day.

Based on the suggestion of \cite{gebbran2021fair}, the maximum capacity of PV inverters $\overline{s}$ is set to the PV generation without PV curtailment. The maximum reactive power output $\overline{q}$ is set to $10\%$ of $\overline{s}$. The values of $V_1$ to $V_4$ are set to $0.94$ p.u, $0.98$ p.u, $1.06$ p.u, and $1.1$ p.u.

According to \cite{zhong2018topology}, we set the voltage at node $0$ as $240$ V and set it as $1$ p.u. For the active power injected into node $1$, it is the sum of the loads for all households at $t$ moment. For the reactive power injected into node $1$, which is $q^{br}_{1,t}$, we set it as $0$. The reason is that $q^{br}_{h,t}$ means that the PV inverter at node $h-1$ injects reactive power during PV curtailment, while there is no PV inverter at node $0$. 

\begin{figure}[t]
    \centering
    \includegraphics[width=\linewidth]{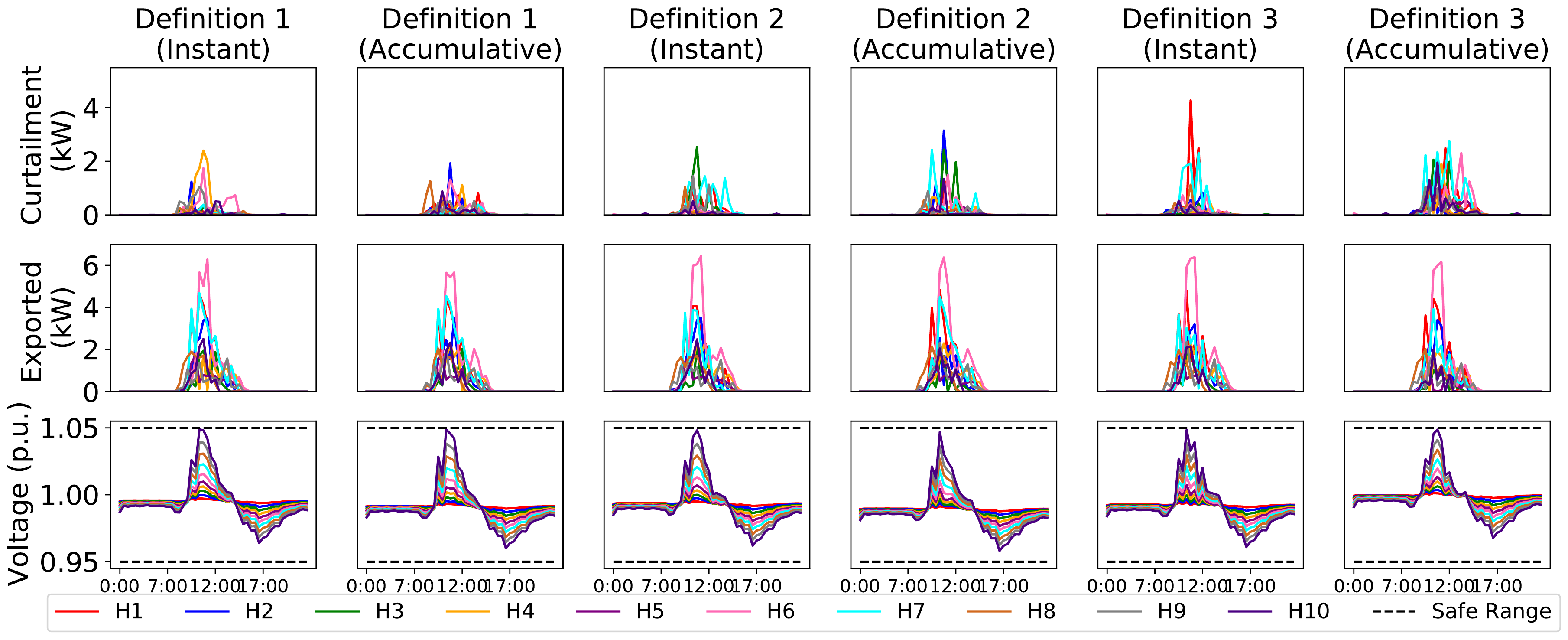}
    \caption{PV curtailment under different fairness definitions.}
    \vspace{-4mm}
    \label{f3}
\end{figure}

Due to the reverse power flow caused by households selling excess power to the grid during PV generation peaks, if a household gets closer to the end of the feeder, its voltage will rise more. Thus, we trained the SAC model to achieve fair PV curtailment based on different fairness definitions, and the results are shown in Fig. \ref{f3},
with more detailed heatmaps of the curtailment in Appendix \ref{ap4}. We observe that the controller limits voltage to the safe range. Comparing across the instant and accumulative cases, we observe that instant curtailment is concentrated in the peak PV generation period, while the PV curtailment in accumulative cases is spread out. This was expected, because the instant case has more stringent per-timestep fairness requirements, while the other case is more concerned with total daily PV curtailment.

Comparing the three definitions, in both the Instant Case and Accumulative Case, the mean of total PV curtailments for each household in the Definition $1$ is slightly smaller than others. The reason for this result is that in the Definition $1$ scenario, the maximum PV curtailment can directly influence the reward, while other scenarios do not have this setting. In addition, because Definition $3$ cares about exports instead of directly limiting PV curtailment value or ratio, the mean of total PV curtailments in the Definition $3$ is the most in these definitions. More specifically, in the Instant Case, the means of total PV curtailments for Definition $1 - 3$ are $3.25$kW, $3.64$kW, and $3.75$kW. In the Accumulative Case, the means of total PV curtailments for Definition $1 - 3$ are $3.02$kW, $4.11$kW, and $7.09$kW. Due to different concerns of different fairness definitions, the agent chooses different policies to ensure that the voltage is within a safe range and no household is subject to unfair curtailments.

\begin{figure}[!t]
\centering
\includegraphics[width=1.0\linewidth]{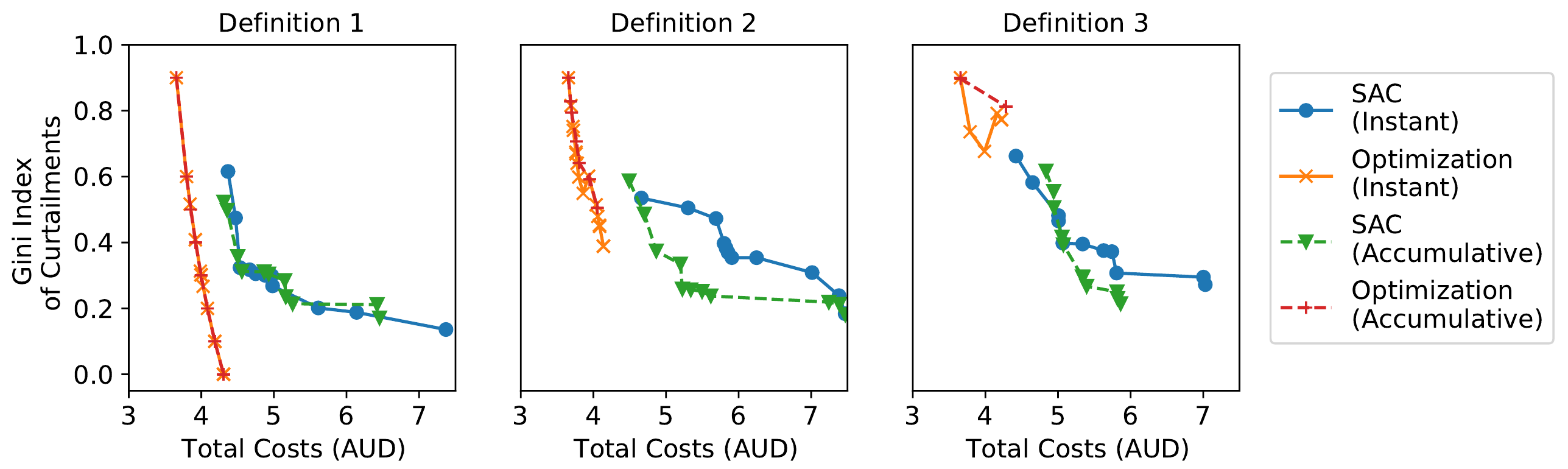}
\caption{Pareto curves of SAC and optimization model.}\label{pareto}
\vspace{-3mm}
\end{figure}

We plot the Pareto curve of the SAC model in Fig.~\ref{pareto} to explore how the model trade-offs between minimizing power cost and improving fairness. We compare our results with optimization assuming known parameters~\cite{gebbran2021fair}, which bounds the RL agent's possible performance. The optimization problem is to minimize the cost of electricity while ensuring safety and improving fairness. The optimization model formulation is presented in Appendix \ref{ap2}.

In Fig.~\ref{pareto}, the $x$-axis represents the total cost of electricity for all households, and the $y$-axis represents the Gini Index of the total PV curtailment for households, which smaller values indicating better fairness. The decrease in Gini Index leads to a rise in electricity costs because a fair curtailment solution can result in some additional PV curtailment \cite{gebbran2021fair}. Since instant cases have fairness requirements for each moment, but accumulative cases are more concerned with the overall situation of the day, accumulative cases are usually more flexible. This means that accumulative cases can produce better results. Our model is slightly inferior to the optimization model in terms of saving electricity costs and improving the fairness of Definition $1$. This result is because the optimization model uses all parameters of the entire distribution network to calculate the optimal solution. In contrast, the SAC model agent can observe only the parameters in the state. Our model performs better for reducing the Gini Index for Definition $2$ and $3$ because optimization models in these definitions do not target the amount of PV curtailment directly. In summary, the most crucial advantage of our model is that it does not need to observe all the distribution network information, and it still achieves a good effect in reducing the Gini Index, which means that it can effectively improve fairness in practice.

\section{Conclusion}\label{6}
In conclusion, we developed an RL-based model to solve the fair curtailment problem. We first described this problem, which means minimizing electricity costs while ensuring that the voltage is in a safe range and improving fairness. We modeled a distribution network to explore the impact of PV curtailment on voltage and household interactions. We also specified several different definitions of fairness. Finally, we developed a SAC model and obtained preliminary experimental results.

Our vision is to build a model that can solve fairness problems in real-world scenarios. Therefore, we will consider reactive power control in the future and employ various fairness measures to obtain a more balanced view. We will also investigate how to select one of the three fairness definitions for a particular scenario. In addition, we will gradually extend our model to consider a longer control period, more households, and more complex network topologies. Moreover, phase imbalance and high-voltage grid fluctuations also need to be considered in our future work.

\section{Acknowledgements}\label{7}
We would like to thank anonymous reviewers for their constructive comments. 
This work has been supported in part by the Australian Research Council (ARC) Discovery Early Career Researcher Award (DECRA) under Grant DE230100046 and the Monash Energy Institute Professional development awards for ECR Women in Energy.

\clearpage
\bibliographystyle{ACM-Reference-Format}
\bibliography{main.bib}
\clearpage
\appendix
\section{Appendices}
\subsection{Illustrated fairness definitions}\label{ap1}
\begin{figure}[!htbp]
    \centering
    \includegraphics[width=0.7\linewidth]{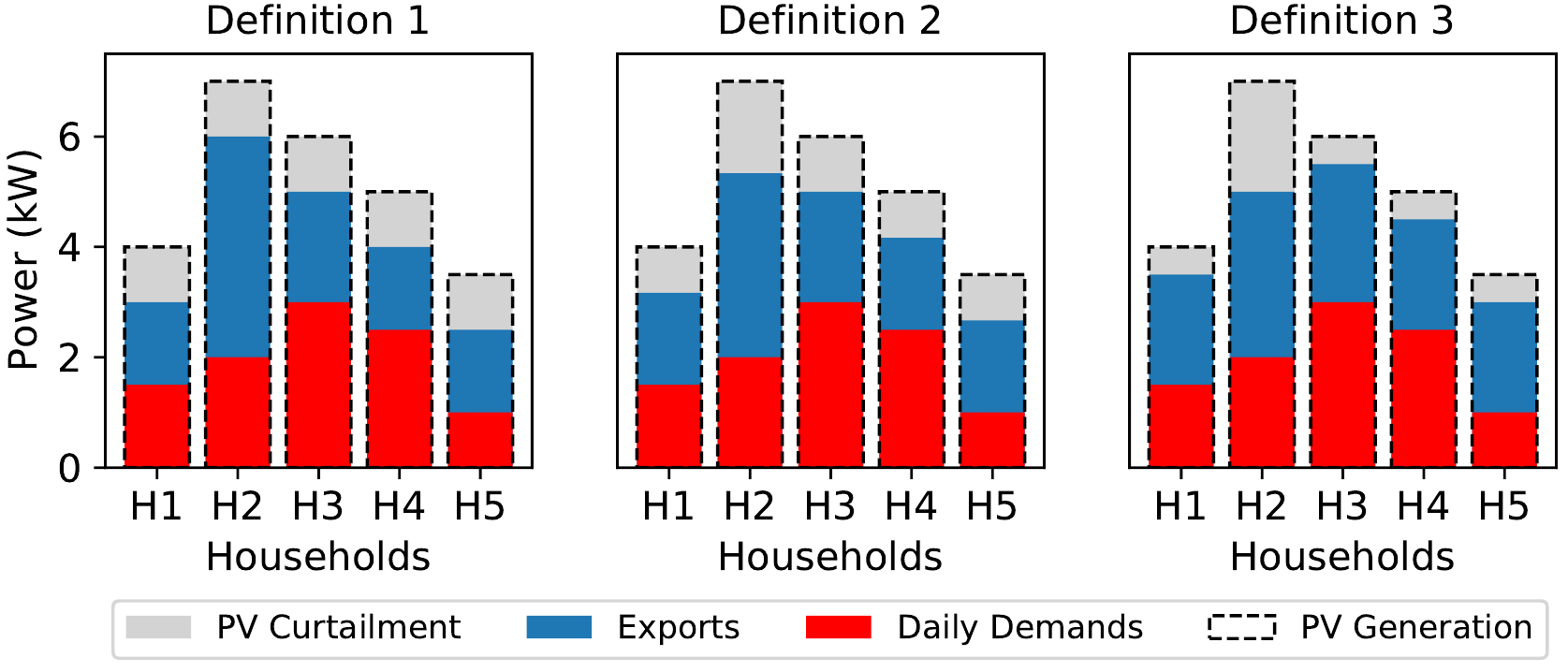}
    \caption{Differences between fairness definitions.}
    \label{diff}
\end{figure}

Fig. \ref{diff} illustrates the differences between these definitions.

In Fig. \ref{diff}, the dashed boxes represent the total amount of PV generation, which is split between curtailed (grey) and consumed (red). The remainder of the generation (blue) is sold to the grid. 

\subsection{Optimization model}\label{ap2}
For the optimization model, we aim to minimize the cost of electricity while ensuring safety and improving fairness. We modeled this optimization problem as a linear programming problem and used the MiniZinc solver. Our objective function includes the cost of electricity consumption and the penalty due to unfair curtailments. Thus, we can tradeoff the cost minimization and fairness maximization by the weight $\beta$ in this objective function. The optimization problem is formulated as
\begin{equation}
\begin{aligned}
\text{min } & \sum_{h \in \mathcal{H}} \sum_{t \in \mathcal{T}} {c_t \times (p^L_{h,t} - p^G_{h,t} \times (1 - \alpha^C_{h,t})) - \beta f(\boldsymbol{\alpha}^C_t)}, \\
\text{s.t. }
& 0 \leq \alpha^C_t \leq 1,
\\& p_{h,t} = p^G_{h,t} \times (1 - \alpha^C_{h,t}) - p^L_{h,t},
\\& p^{br}_{h+1,t} = p^{br}_{h,t} + p_{h,t},
\\& q^{br}_{h+1,t} = q^{br}_{h,t} + q_{h,t},
\\& v_{h+1,t} = v_{h,t} - (r_{h+1,t}p^{br}_{h+1,t} + x_{h+1,t}q^{br}_{h+1,t})/v_{0,t},
\\& 0.95 \leq v_{h,t}\text{ (p.u)} \leq 1.05,
\\& q_{h,t} = \overline{q} \times \mathbbm{1}(v_{h,t} \leq V_1) + \overline{q}\frac{V_2 - v_{h,t}}{V_2 - V_1} \times \mathbbm{1}(V_1 < v_{h,t} < V_2), \\& - \overline{q}\frac{v_{h,t} - V_3}{V_4 - V_3} \times \mathbbm{1}(V_3 < v_{h,t} < V_4) - \overline{q} \times \mathbbm{1}(v_{h,t} \geq V_4),
\end{aligned}
\end{equation}
where $\beta$ is the weight for fairness penalties.

\subsection{Learning curves}\label{ap3}
\begin{figure}[!htbp] %
    \centering
    \includegraphics[width=0.75\linewidth]{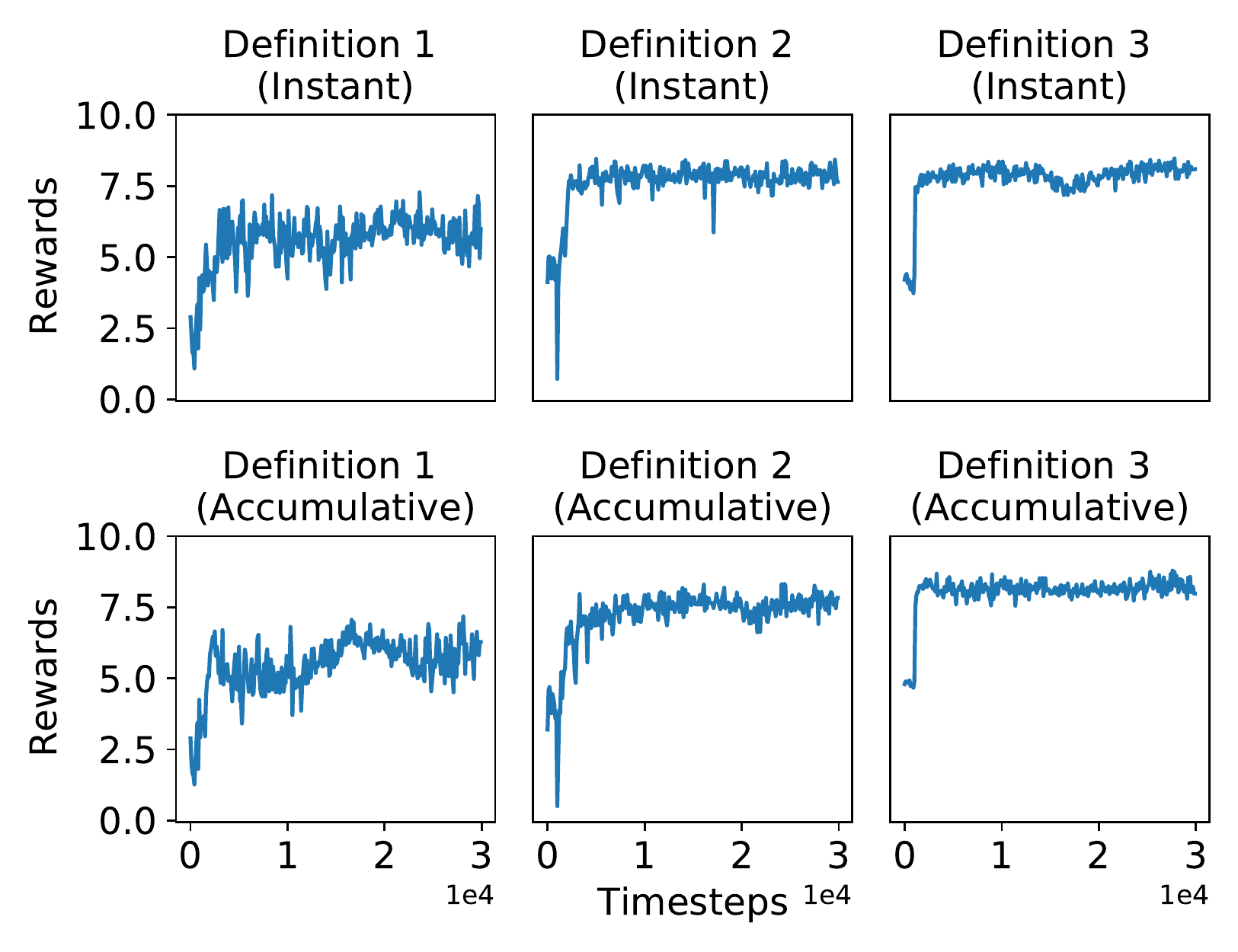}
    \caption{Learning curves for different fairness cases.}
    \label{lc}
\end{figure}

We demonstrate the training process of the SAC model with learning curves, as shown in Fig. \ref{lc}. Each learning curve shows how the per-step reward increases throughout a single run of the training process. The reward is defined in \eqref{eq:reward}, taking different forms according to the type of fairness definitions and instant/accumulative cases. The reward value means the sum of the ratio of PV generation after curtailment to the generation before curtailment for all households, minus the penalties. The maximum reward per step of the policy is $10$, which means no curtailment for each of the $10$ households and no penalties.

\subsection{Learned curtailment policies}\label{ap4}
We complement Fig. \ref{f3} with heat maps to show exported and curtailed PV of households under different fairness cases. Since PV generation is not available at all times of the day, we only selected the time period with the highest PV production (8:00 - 15:00) to plot heat maps.
\clearpage
\begin{figure*}[!htbp] %
    \centering
    \includegraphics[width=0.49\linewidth]{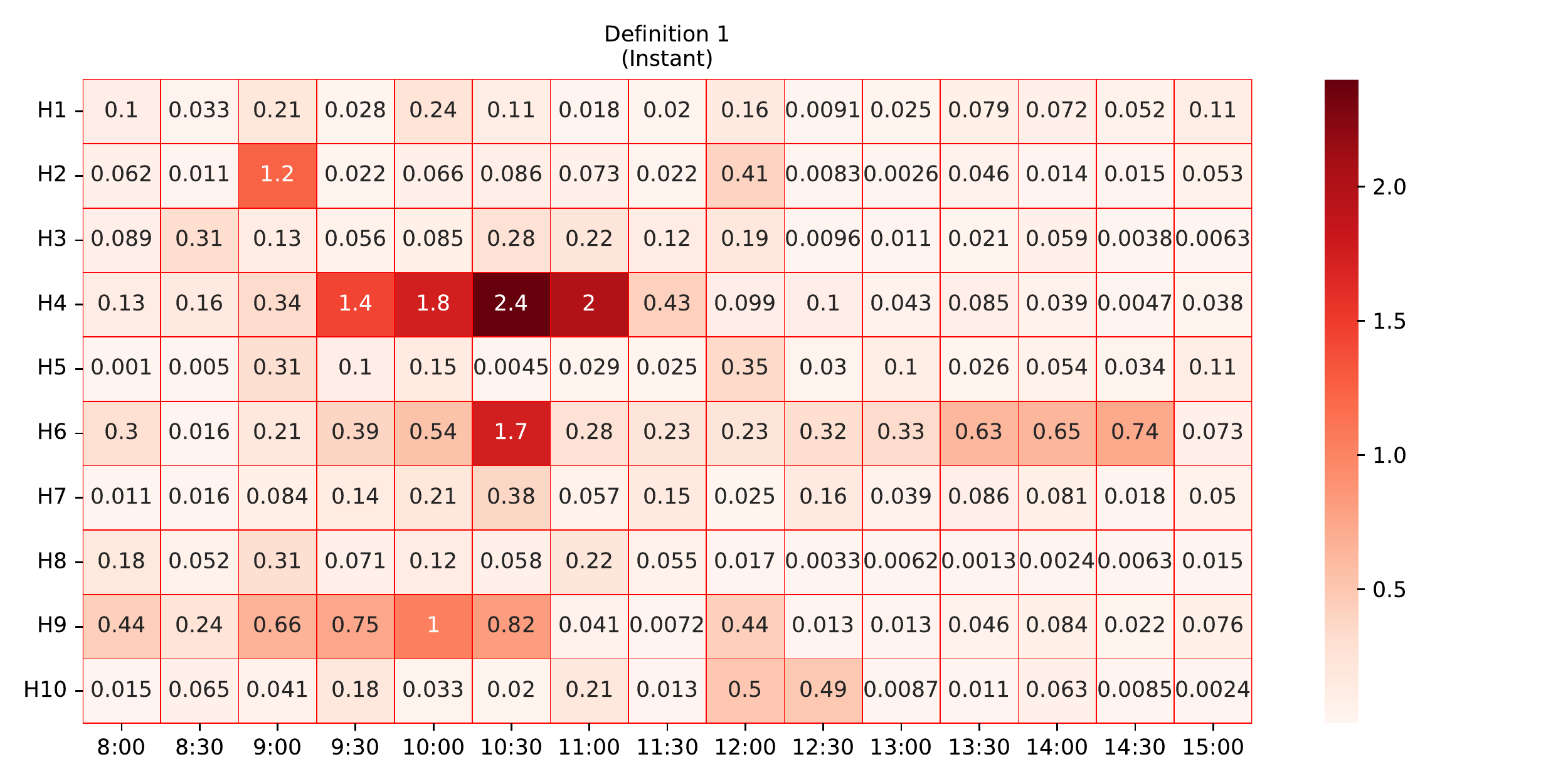}
    \includegraphics[width=0.49\linewidth]{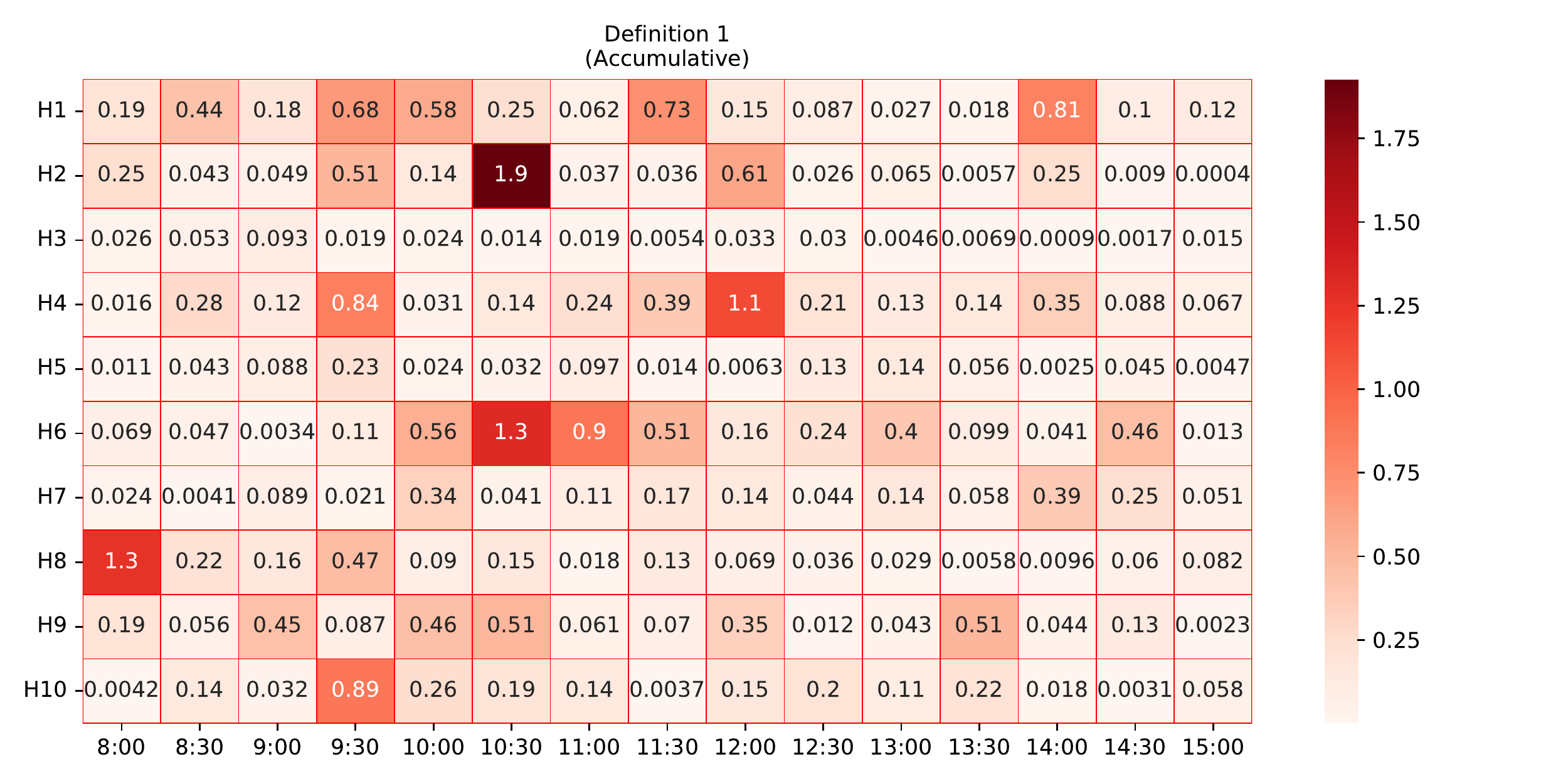}
    \includegraphics[width=0.49\linewidth]{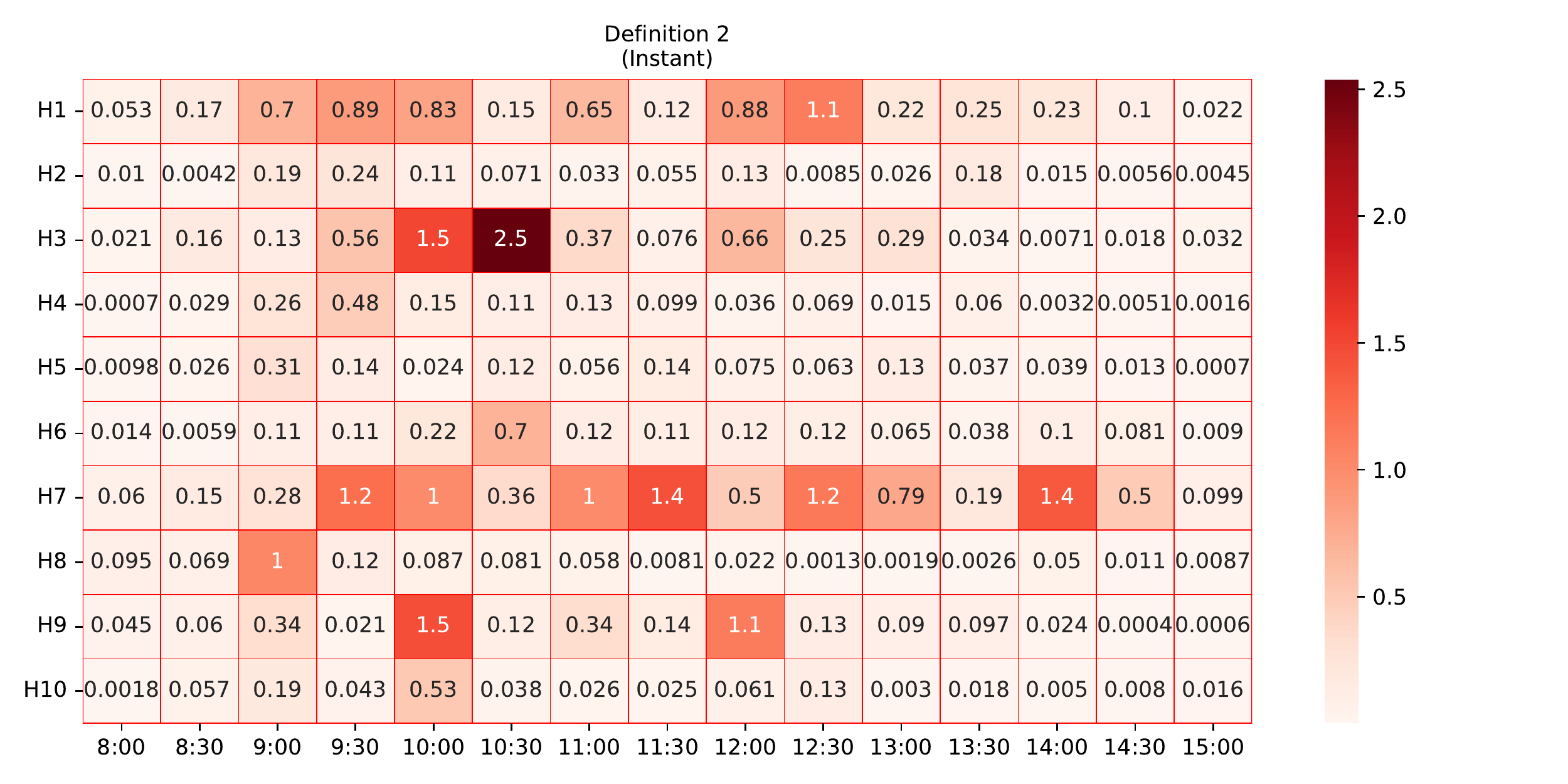}
    \includegraphics[width=0.49\linewidth]{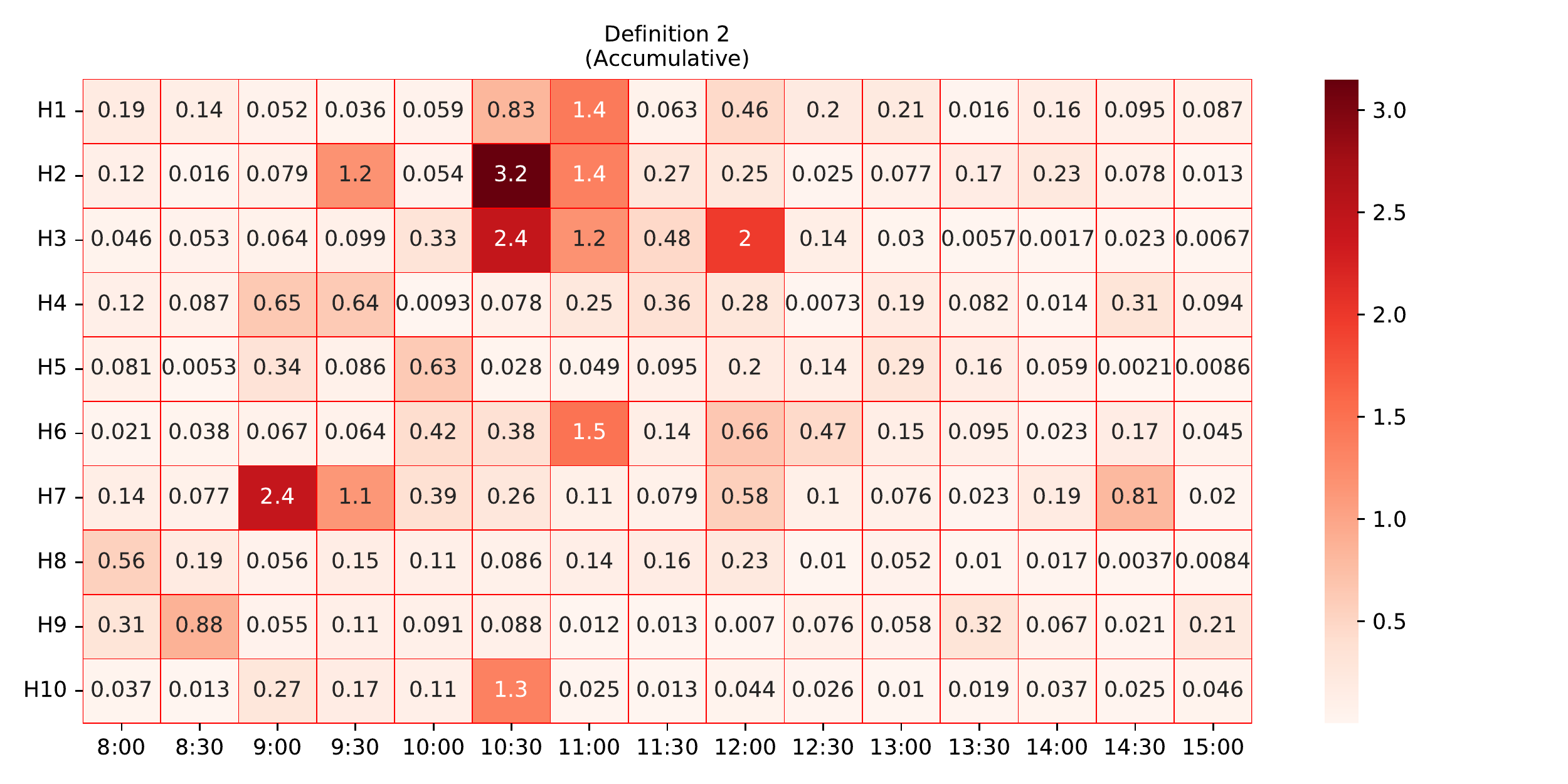}
    \includegraphics[width=0.49\linewidth]{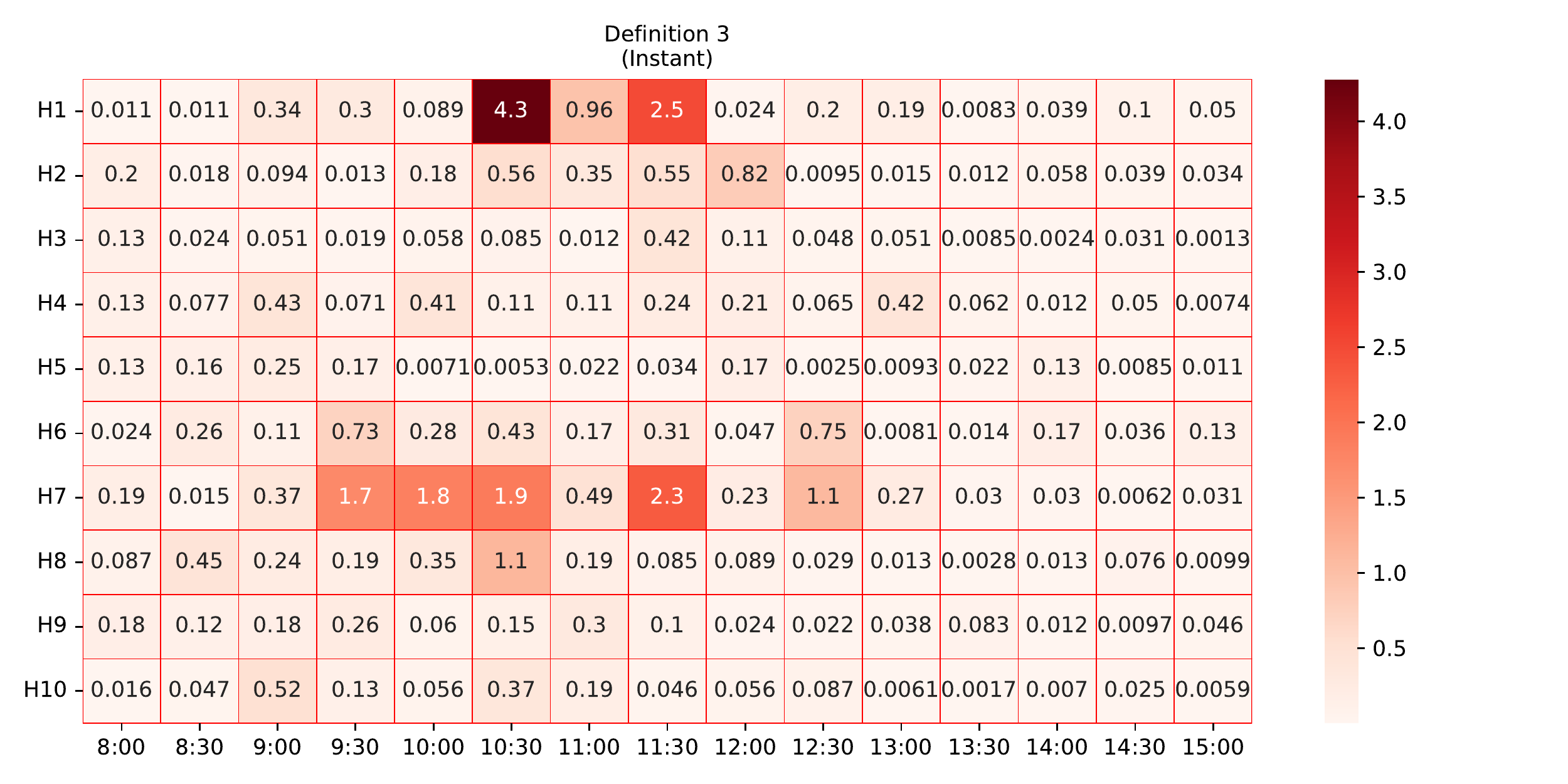}
    \includegraphics[width=0.49\linewidth]{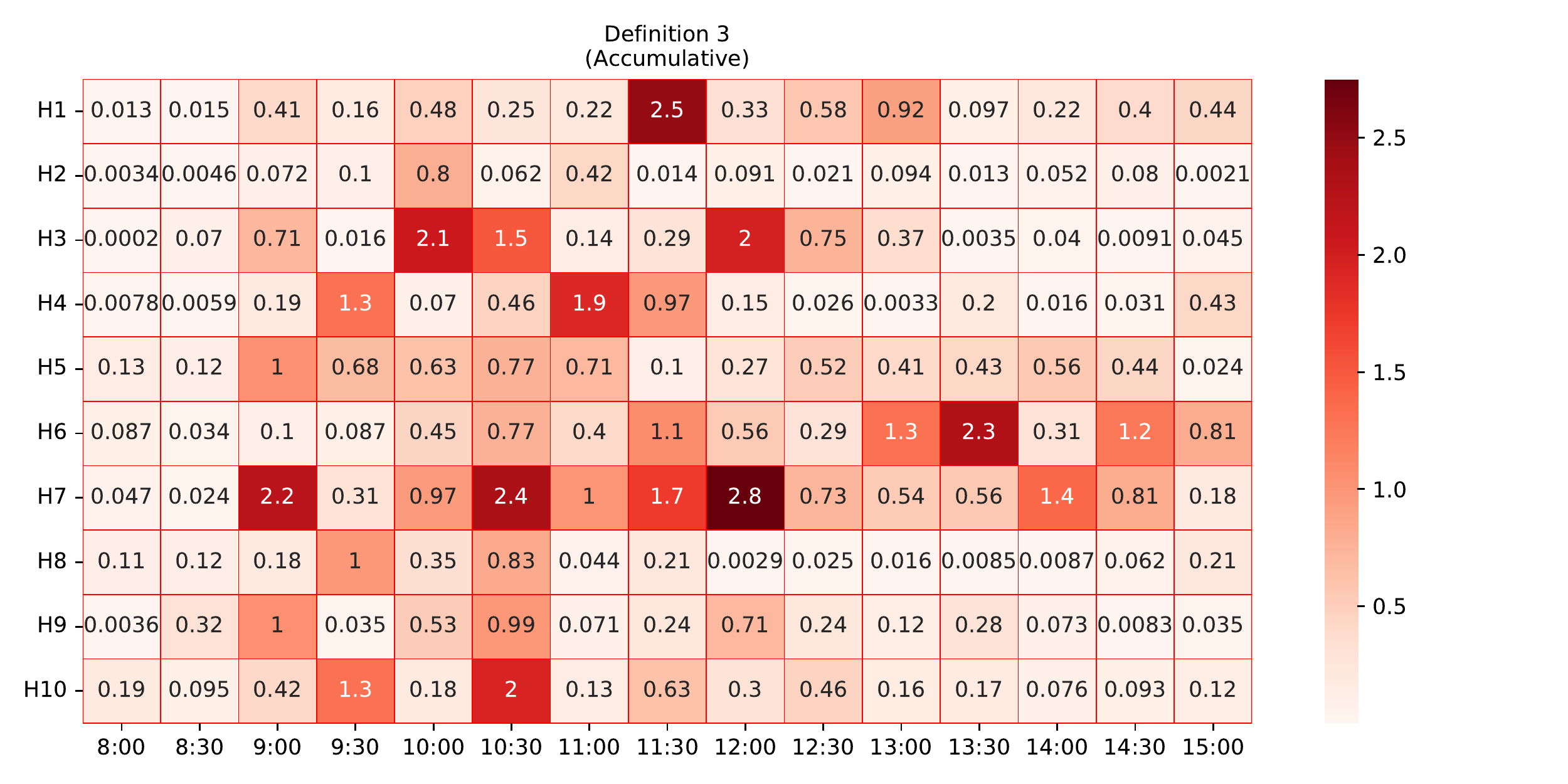}
    \caption{PV curtailments during 8:00 - 15:00.}
    \label{hm}
\end{figure*}

\begin{figure*}[!htbp] %
    \centering
    \includegraphics[width=0.49\linewidth]{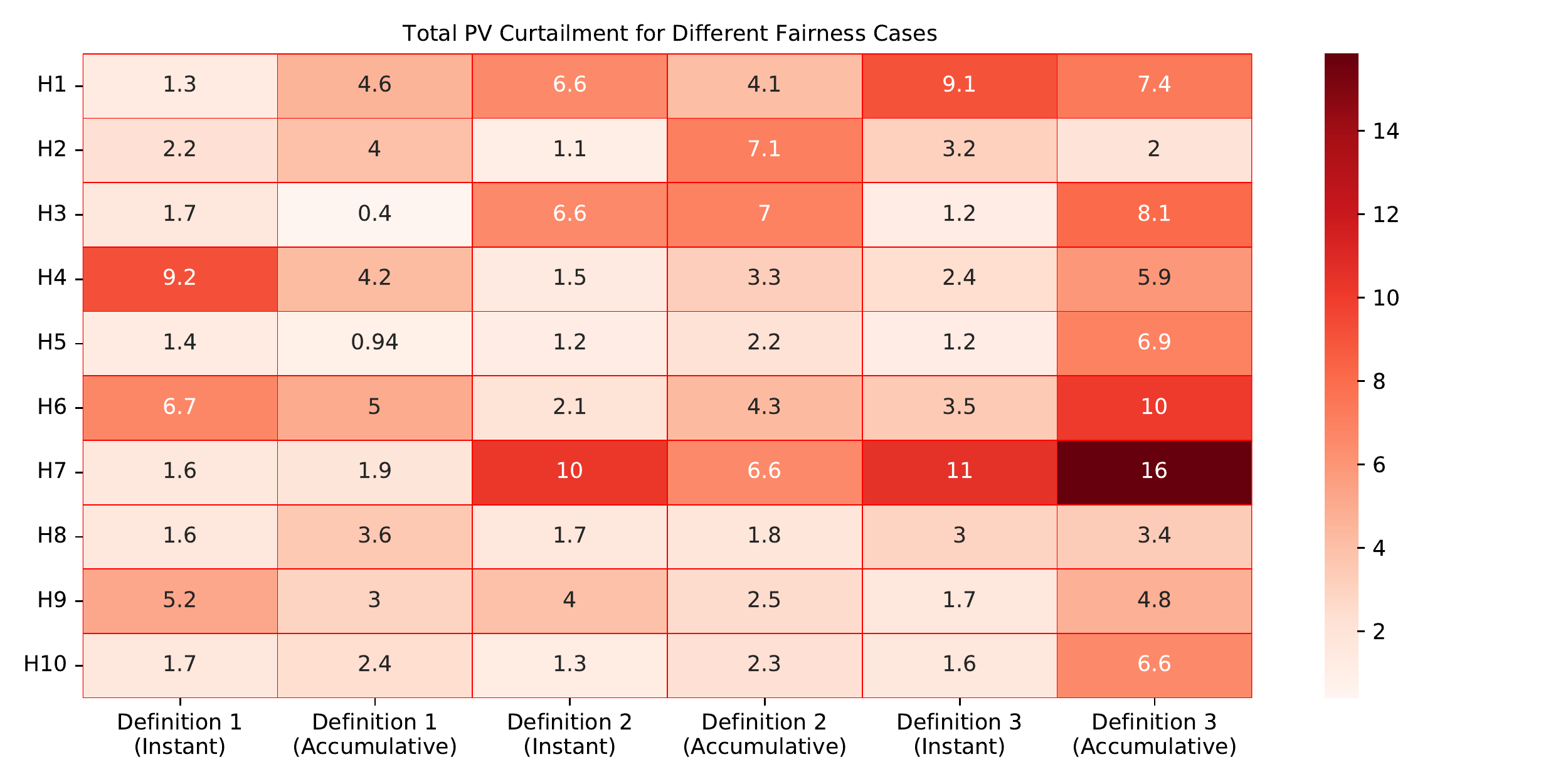}
    \caption{Total PV curtailments for the whole day.}
    \label{total_hm}
\end{figure*}

\begin{figure*}[!htbp] %
    \centering
    \includegraphics[width=0.49\linewidth]{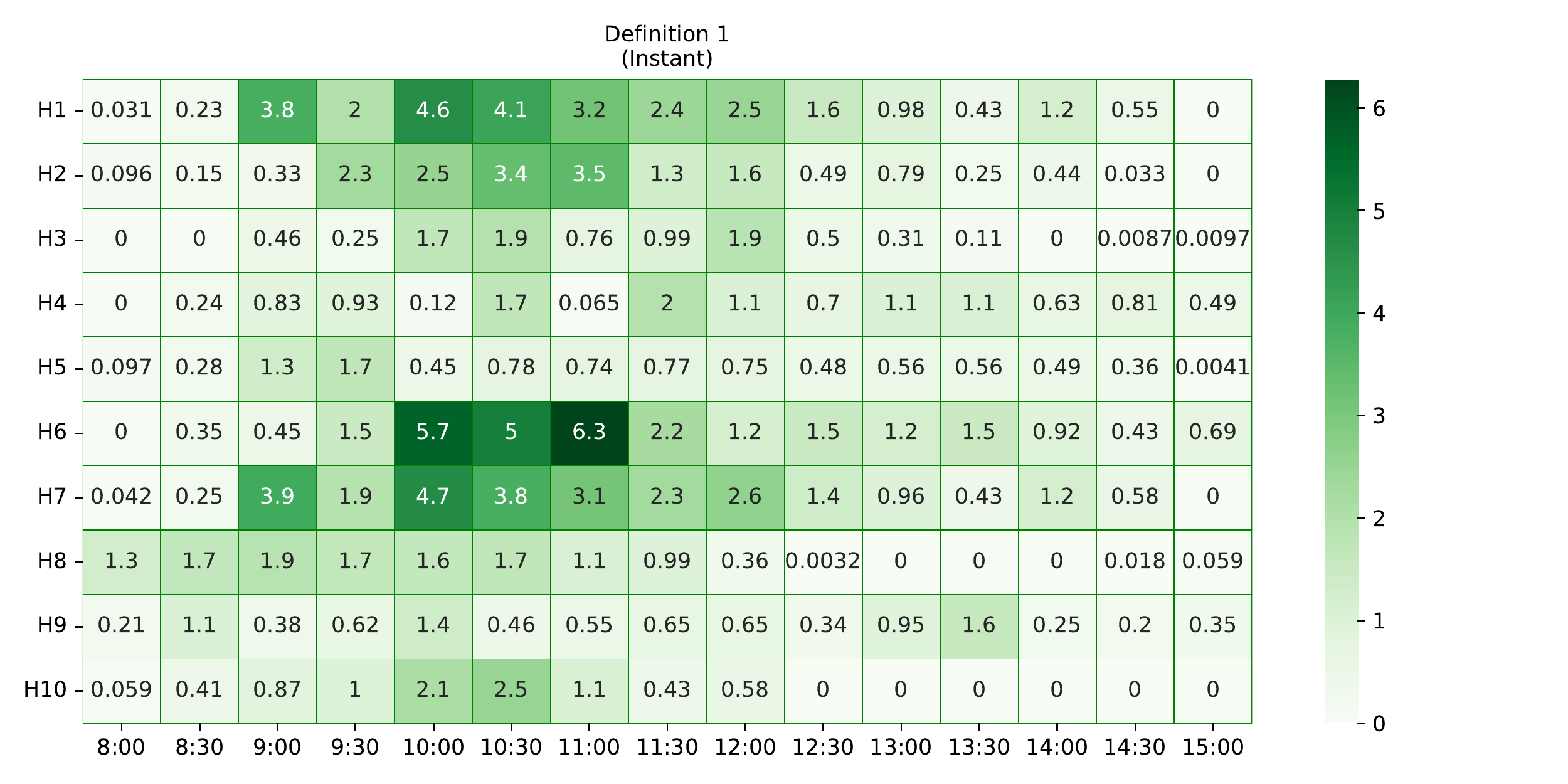}
    \includegraphics[width=0.49\linewidth]{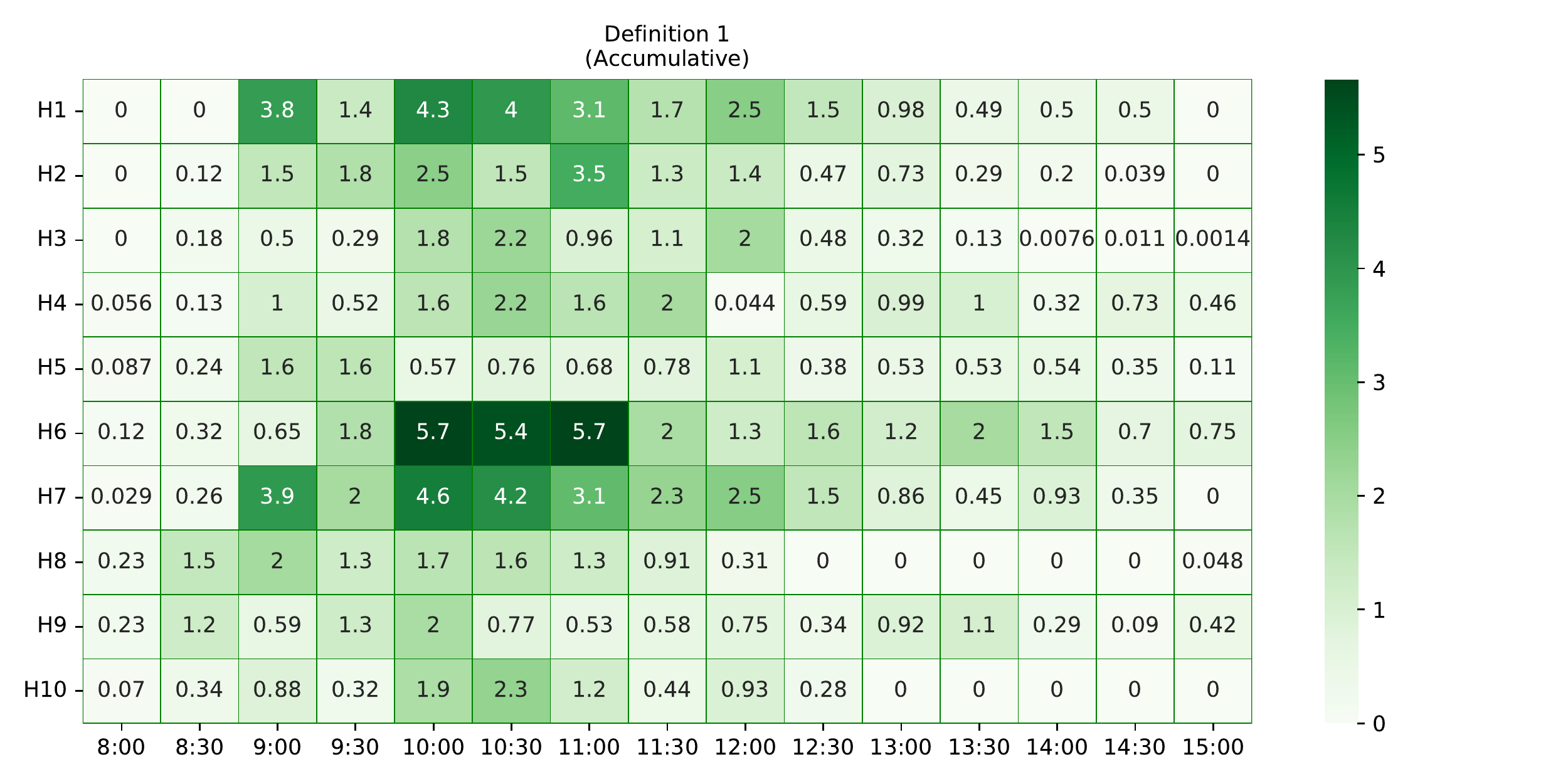}
    \includegraphics[width=0.49\linewidth]{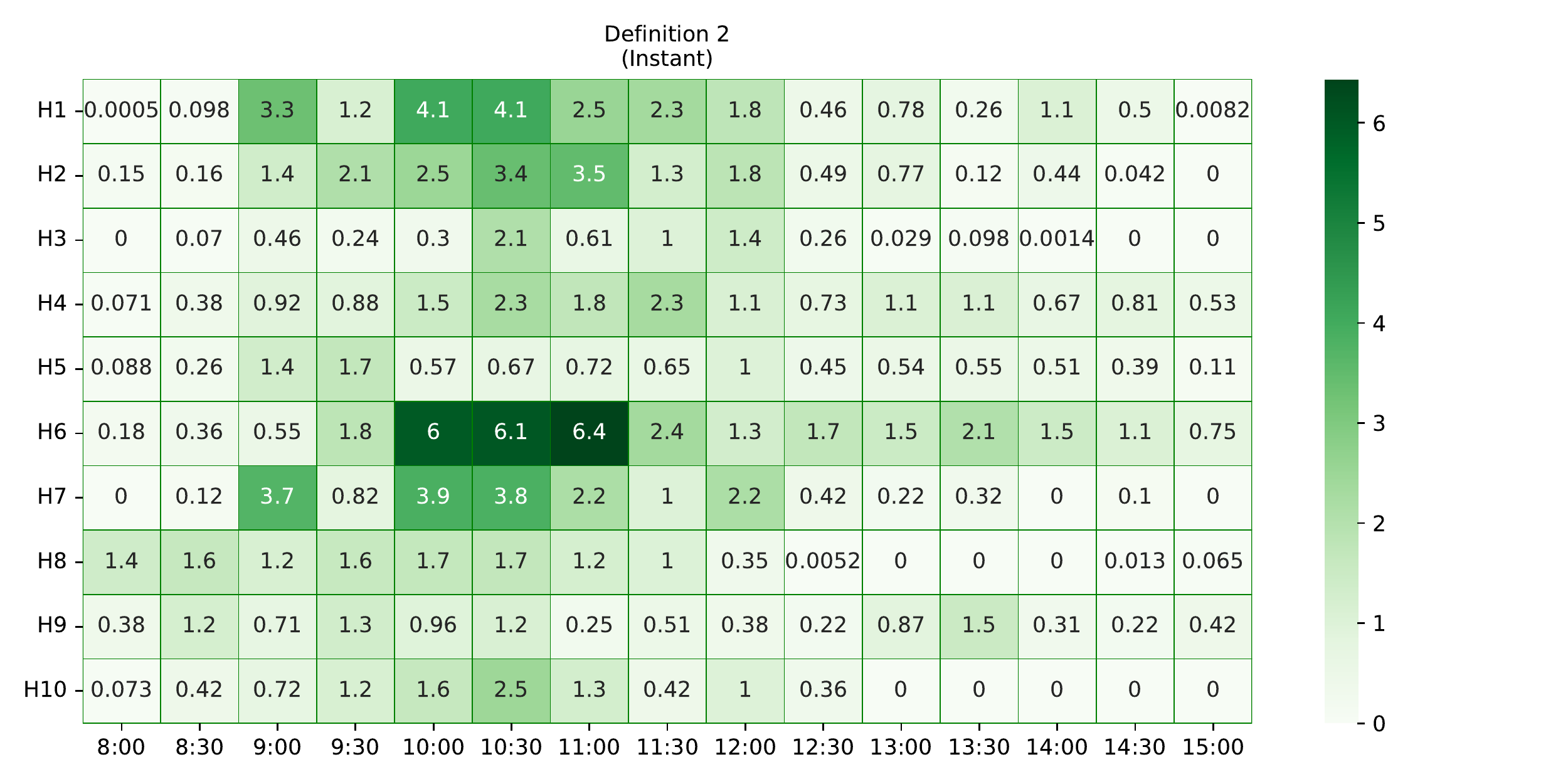}
    \includegraphics[width=0.49\linewidth]{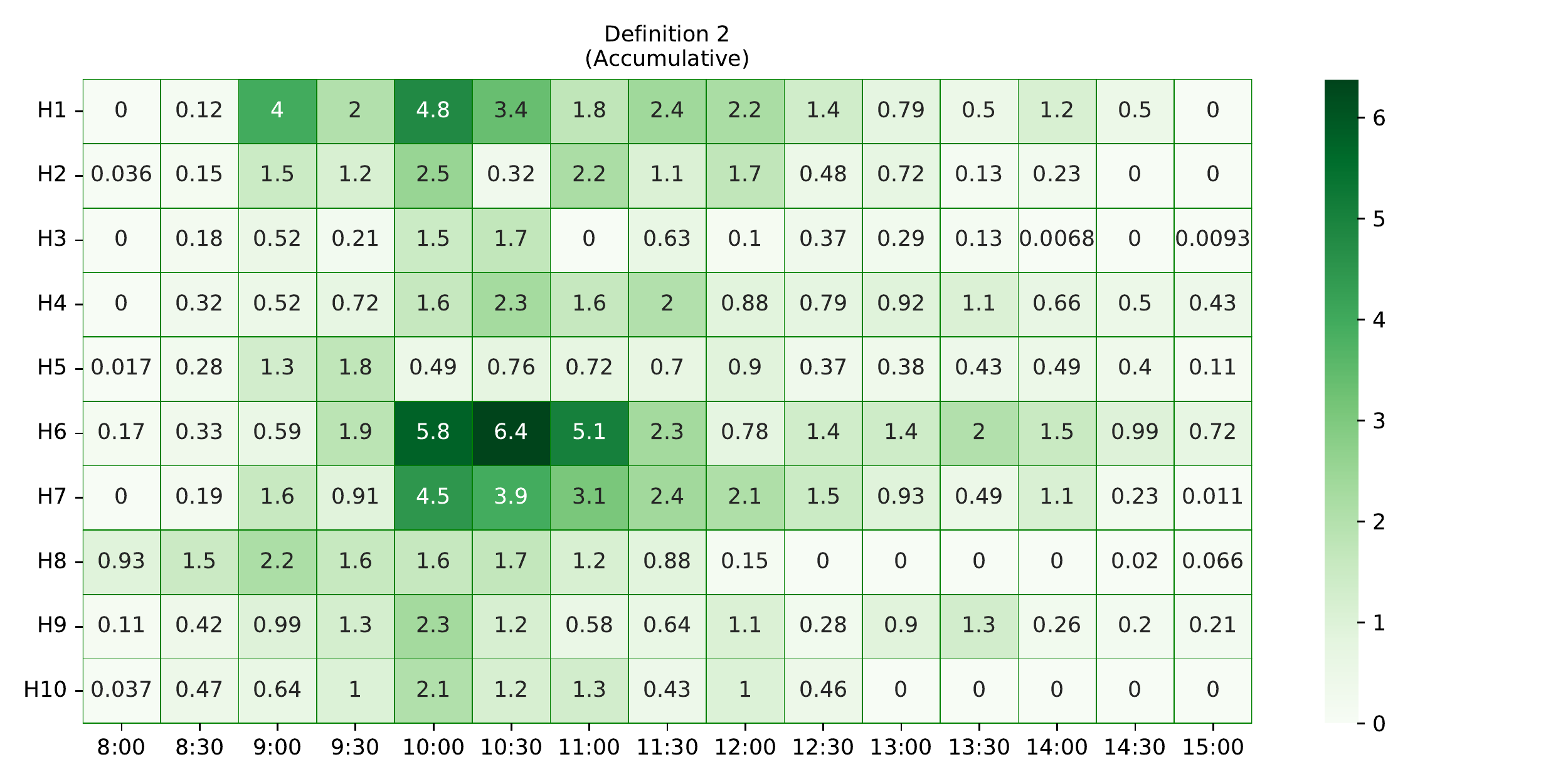}
    \includegraphics[width=0.49\linewidth]{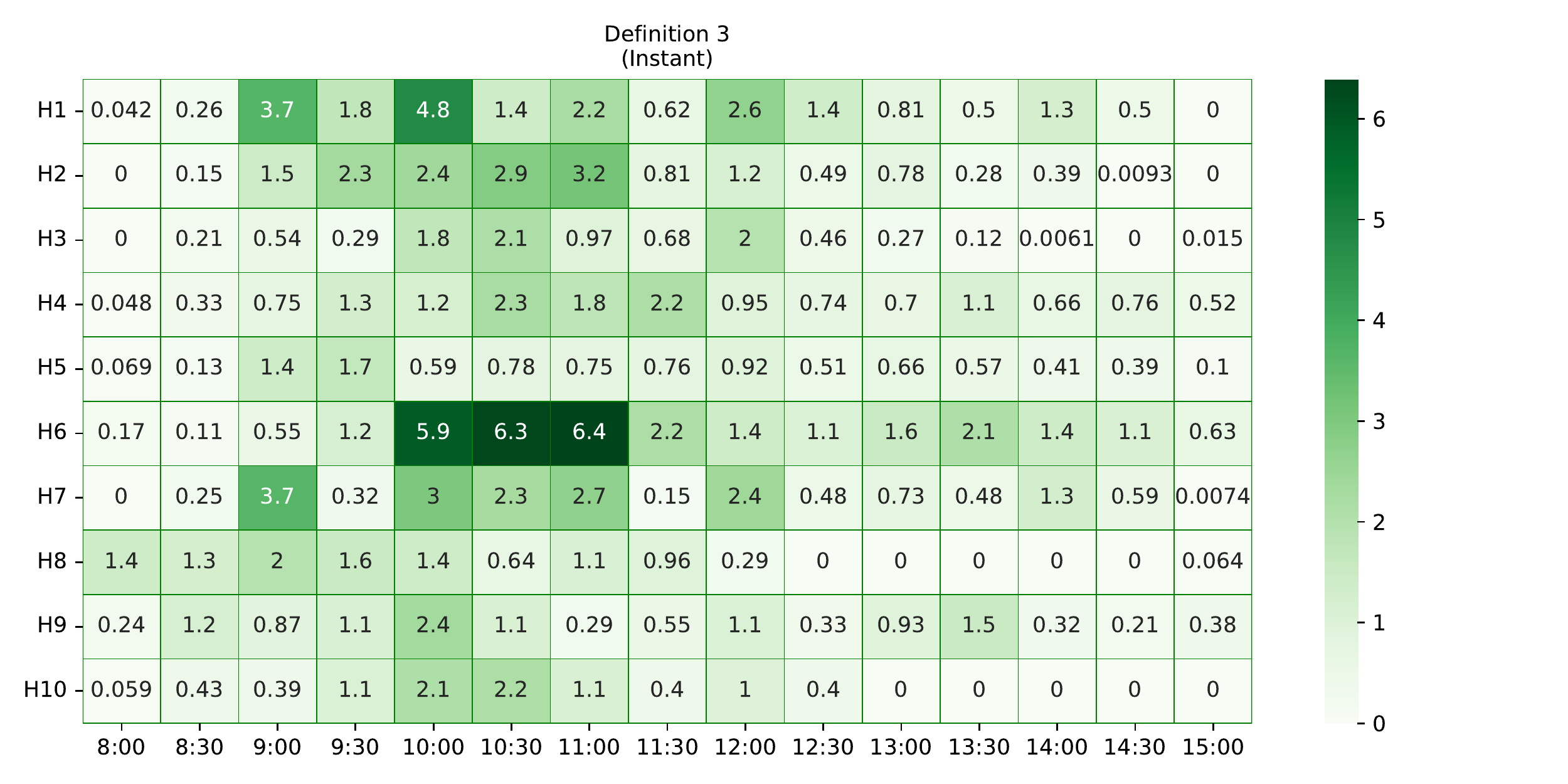}
    \includegraphics[width=0.49\linewidth]{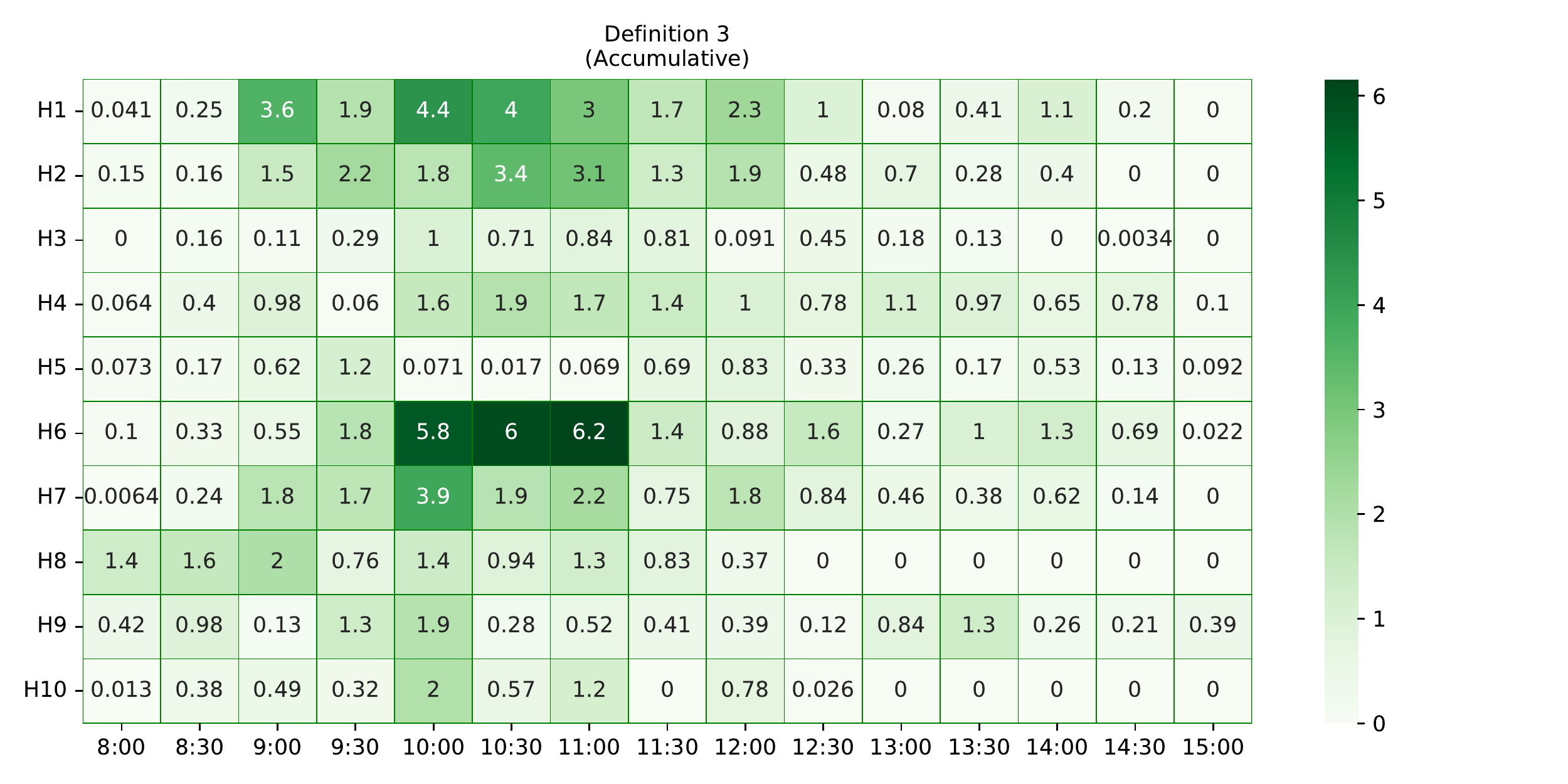}
    \caption{Exported PV during 8:00 - 15:00.}
    \label{hm_sell}
\end{figure*}

\begin{figure*}[!htbp] %
    \centering
    \includegraphics[width=0.49\linewidth]{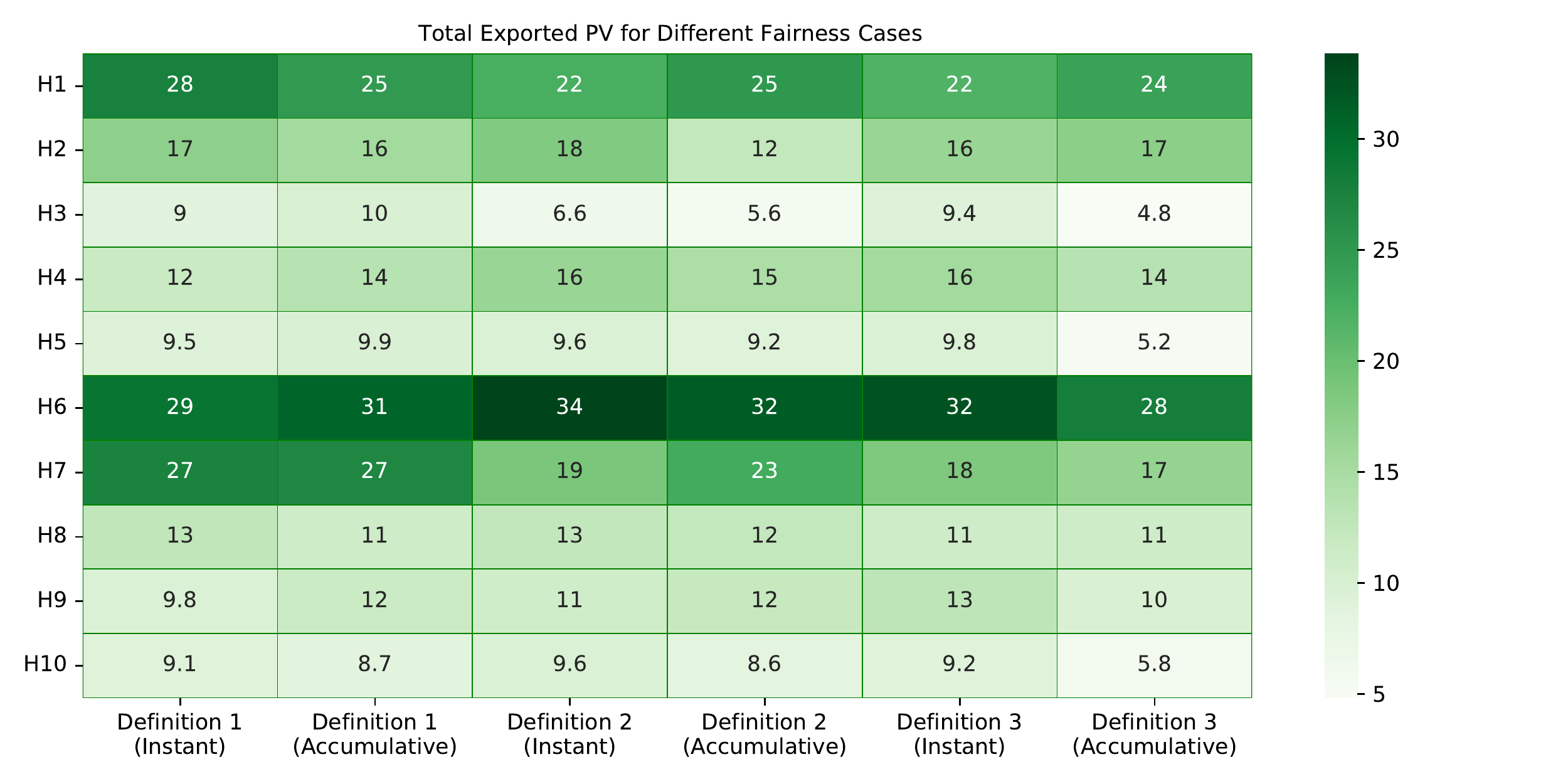}
    \caption{Total exported PV for the whole day.}
    \label{total_hm_sell}
\end{figure*}

\end{document}